\documentclass{jetpl}
\twocolumn

\usepackage{graphicx}

\lat

\title{Cyclic Period-3 Window in Antiferromagnetic Potts and Ising Models on Recursive Lattices}

\rtitle{Cyclic Period-3 Window in Antiferromagnetic \ldots}

\sodtitle{Cyclic Period-3 Window in Antiferromagnetic Potts and Ising Models on Recursive Lattices}

\author{N.\,S.\,Ananikian\/\thanks{ananik@mail.yerphi.am},
L.\,N.\,Ananikyan, L.\,A.\,Chakhmakhchyan}

\rauthor{N.\,S.\,Ananikian, L.\,N.\,Ananikyan, L.\,A.\,Chakhmakhchyan}

\sodauthor{Ananikian}

\address{A.I. Alikhanyan National Science Laboratory, Alikhanian Br. 2, 0036 Yerevan, Armenia}

\abstract{The magnetic properties of the antiferromagnetic Potts model with two-site interaction and the antiferromagnetic Ising model with three-site interaction on recursive lattices have been studied. A cyclic period-3
window has been revealed by the recurrence relation method in the antiferromagnetic $Q$-state Potts model on
the Bethe lattice (at $Q < 2$) and in the antiferromagnetic Ising model with three-site interaction on the Husimi
cactus. The Lyapunov exponents have been calculated, modulated phases and a chaotic regime in the cyclic
period-3 window have been found for one-dimensional rational mappings determined the properties of these
systems.}


\begin{document}

\maketitle

The Potts and Ising models played an important
role in the theories of phase transitions and critical
phenomena \cite{model, ising}. Since the Potts model is not solved
exactly in the general case (solutions have not been
found for dimensions $d > 2$ and for nonzero magnetic
field), various approximations are used. In particular,
the Bethe-Peierls approximation reduces the study of
the system dynamics to analysis of the behavior of
rational mappings obtained by the recurrence relation
method \cite{Ananikyan}. This method can also be applied to the
generalized Bethe lattice (Husimi cactus) \cite{RNA},
to
describe frustrated antiferromagnetic systems with
multisite interaction. Such models simulate the properties of the magnetization of solid $^3\mathrm{He}$ \cite{He3}. At the
same time, this dynamic approach can also be applied
to study gauge systems described by both multidimensional and one-dimensional mappings
\cite{map}. In our case,
recurrence relations describing the models in the presence of antiferromagnetic pairing of the lattice sites
exhibit a quite complex behavior including the doubling cascade, chaos, and cyclic window with period $p= 3, 5, 6, ...$. Such windows were revealed and studied
both theoretically and experimentally in a number of
other systems of applied interest \cite{three}. The Lyapunov
exponent can be considered as an order parameter for
determining the geometric and dynamical properties
of the attractor of such systems \cite{lyapunov}.

The aim of this work is to analyze the cyclic period-3 window for rational mappings describing the antiferromagnetic $Q$-state Potts model on the Bethe lattice
($Q < 2$) and the antiferromagnetic Ising model with
three-site interaction on the Husimi cactus (tree).
This window is represented by the laminar phase with
an incorporated chaotic behavior. The transition from
the chaotic regime to the period-3 regime occurring through tangent bifurcation (type-I intermittency)
\cite{Pomeau}, as well as subsequence doubling of the period
through the type-II intermittency (doubling bifurcation), is considered.

In the presence of external magnetic field, the $Q$-state Potts model on the Bethe lattice \cite{model} is specified
by the Hamiltonian

\begin{eqnarray}
{\mathcal{H}}=-J\sum_{(i,j)}\delta({{\sigma_{i},\sigma_{j}} })-H
\sum_{i}\delta({{\sigma_{i},Q} }), \label{1}
\end{eqnarray}
where $\sigma_i=1, 2, ..., Q$, $\delta(x, y)$ is the Kronecker delta;
the first and second sums are taken over all of the edges
and sites of the lattice, respectively; and $J>0$ corresponds to antiferromagnetic pairing. The partition
function and magnetization at the central site can be
represented as
\begin{gather}
\mathcal{Z}=\sum_{\{{\sigma}\}}{e^{-\frac{\mathcal{H}}{k_B T}}},\label{2}
\\ M=\langle\delta(\sigma _0,Q)\rangle=\mathcal{Z}^{-1}\sum_{\{\sigma\}}{\delta(\sigma _0,Q)}e^{-\frac{\mathcal{H}}{k_BT}}, \label{3}
\end{gather}
where $k_B$ is the Boltzmann constant (below, we set
$k_B = 1$). Cutting the Bethe lattice at the central site
into $\gamma$ identical branches ($\gamma$ is the coordination number), we represent the partition function in the form
\begin{eqnarray}
\mathcal{Z}=\sum_{\{{\sigma} _0\}}exp\{\frac{H}{T}\cdot\delta(\sigma _0,
Q)\}[g_{n}(\sigma _0)]^\gamma, \label{4}
\end{eqnarray}
where $\sigma_0$ is the central spin and $g_{n}(\sigma _0)$ is the contribution of each of the $\gamma$ identical branches. Following a
known procedure described in \cite{Ananikyan, He3}, we obtain
\begin{gather}
x_{n}=f_1(x_{n-1}), \nonumber
\\ f_1(x)=\frac{e^{\frac{H}{T}}+(e^{\frac{J}{T}}+Q-2)x^{\gamma-1}}{e^{\frac{H+J}{T}}+(Q-1)x^{\gamma-1}}, \label{7}
\end{gather}
where $x_n=g_n(\sigma\neq Q)/g_n(\sigma=Q)$.
The rational mapping given by Eq.~(\ref{7}) is known as the Potts-Bethe mapping.
Taking into account Eq.~(\ref{3}), we obtain the following
expression for the magnetization:
\begin{eqnarray}
M_n=\langle\delta(\sigma_0, Q)\rangle
=\frac{e^\frac{H}{T}}{e^\frac{H}{T}+(Q-1)x_{n}^\gamma}.\label{8}
\end{eqnarray}

\begin{figure*}
\begin{center}
\includegraphics[width=17cm]{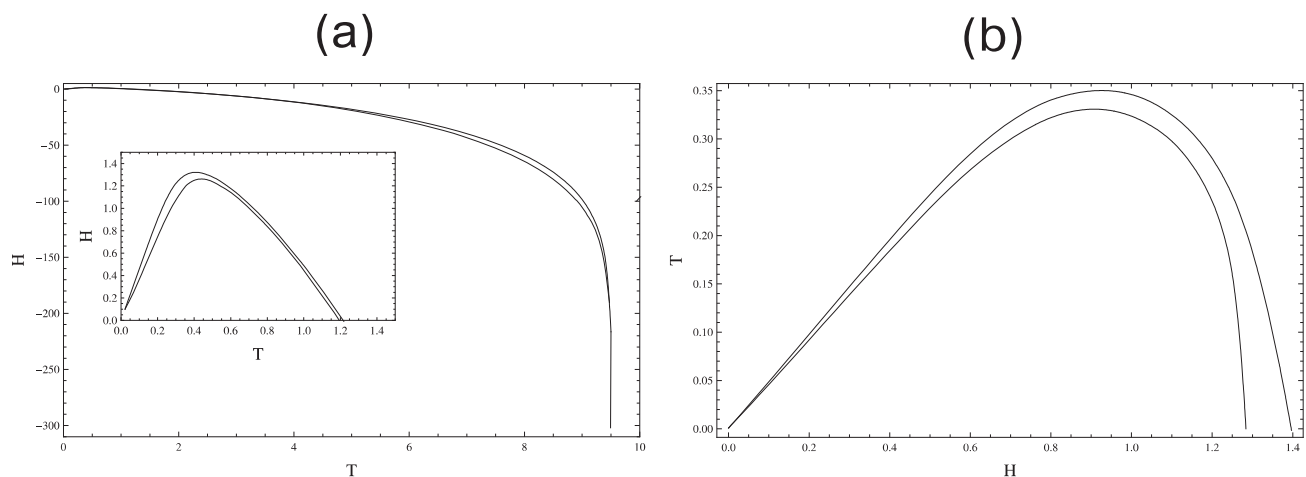}
\caption{\small{Fig. 1. Modulated period-3 phase for (a) the antiferromagnetic Potts model at $Q = 1.1$, $J = –1$, and $\gamma = 3$ (inset shows
the region $H > 0$ at a magnified scale) and (b) the antiferromagnetic Ising model with three-site interaction at $J_3 = -1$ and $\gamma = 3$.
\label{chain}}}
\end{center}
\end{figure*}

The antiferromagnetic Ising model with three-site
interaction on the Husimi cactus in the presence of the
external magnetic field \cite{RNA} is specified by the Hamiltonian
\begin{eqnarray}
{\mathcal{H}}=-J_3\sum_{\bigtriangleup}\sigma_{i}\sigma_{j}\sigma_{k}-H\sum_{i}\sigma_{i}, \label{8.1}
\end{eqnarray}
where $\sigma_i=\pm1$, the first sum is taken over all of the triangles, and second sum is calculated over all of the
sites of the cactus, and $J_3 < 0$. In this case, the rational
mapping and magnetization at the site of the central
triangle are given by the expressions:

\begin{gather}
x_{n}=f_2(x_{n-1}), \nonumber
\\ f_2(x)=\frac{x^{2 (\gamma -1)} e^{\frac{4 H+2 J_3}{T}}+2 e^{\frac{2 H}{T}}
   x^{\gamma -1}+e^{\frac{2 J_3}{T}}}{2 x^{\gamma -1} e^{\frac{2 H+2
   J_3}{T}}+e^{\frac{4 H}{T}} x^{2 (\gamma -1)}+1}, \label{8.2}
\\ M_n=\langle\sigma_0\rangle=\frac{e^{\frac{2 H}{T}} x_n^{\gamma }-1}{e^{\frac{2 H}{T}} x_n^{\gamma }+1}, \label{8.3}
\end{gather}
where $x_n=j_n(\sigma_0=+1)/j_n(\sigma_0=-1)$ ($j_n$ is the contribution from each of $\gamma$ identical branches after cutting in
the central triangle).

Thus, the $Q$-state Potts model on the Bethe lattice
and the antiferromagnetic Ising model with three-site
interaction on the Husimi cactus can be considered as
nonlinear dynamical systems described by rational
mappings. It is of interest to analyze the dependence
of these rational mappings on both parameter $T$ (temperature) and parameter $H$ (external field) at a fixed
coupling constant ($J$ or $J_3$), coordination number $\gamma$,
and the number of states $Q$. The bifurcation points of
the mappings correspond to the phase transition
points with a change in symmetry.

We study the transition from the chaotic regime to
the cyclic period-3 regime through tangent bifurcation
\cite{Pomeau}. Certain values of the temperature $T$ and magnetic
field $H$ specify a curve separating the chaotic and
period-3 regimes (the mapping in the latter regime has
three stable stationary points). In this curve, tangent
bifurcation occurs under the condition
\begin{eqnarray}
\left\{\begin{array}{ll}
f_i^{(3)}(x)=x & \\
(f_i^{(3)}(x))'=1, (i=1, 2), &
\end{array} \right.\label{50}
\end{eqnarray}
where $f^{(3)}(x)=f(f(f(x)))$. Subsequent bifurcations,
responsible for the appearance of a stable cycle with a
period of $3\times2^n$ ($n=2, 3, ...$), correspond to the doubling of the period. For this reason, the next doubling
bifurcation point can be found from the condition
\begin{eqnarray}
\left\{\begin{array}{ll}
f_i^{(3)}(x)=x & \\
(f_i^{(3)}(x))'=-1, (i=1, 2). &
\end{array} \right.\label{51}
\end{eqnarray}

\begin{figure*}
\begin{center}
\includegraphics[width=17cm]{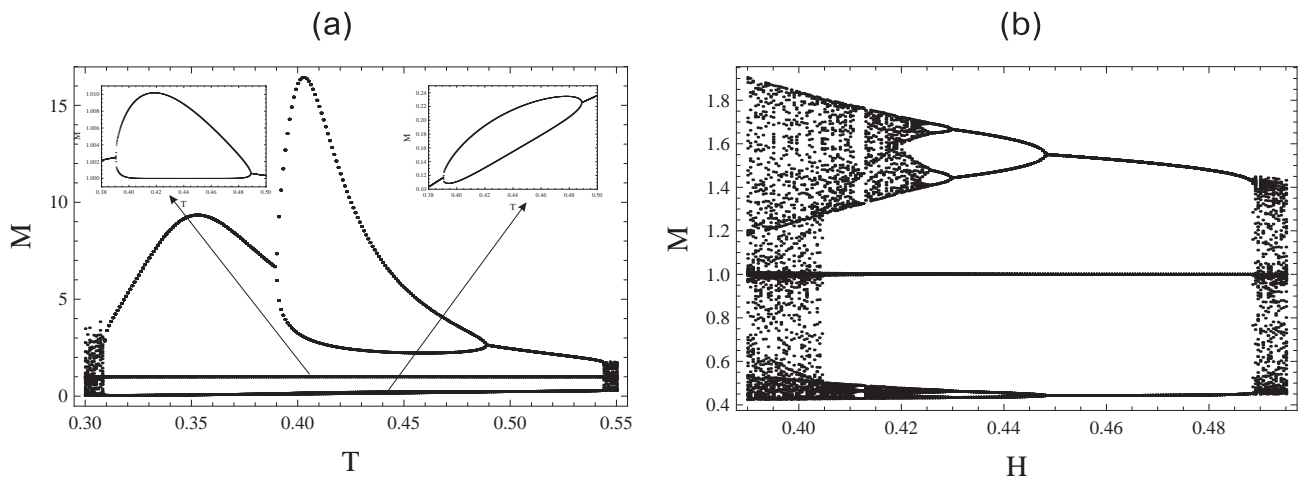}
\caption{\small{Fig. 2. Magnetization in the cyclic period-3 window for
the Potts model at $Q = 1.1$, $J = –1$, and $\gamma = 3$ (a) versus the
temperature for $H = 1.24$ (inset shows the details of the
modulated period-six phase) and (b) versus the magnetic
field for $T = 1$.
\label{chain}}}
\end{center}
\end{figure*}

The region bounded by curves found from conditions (\ref{50}) and (\ref{51})
corresponds to the modulated
period-3 phase 3M0 (i.e., ${\bf 3}\times2^{\bf 0}=3$) of the Potts
model on the Bethe lattice at $i = 1$ and the antiferromagnetic Ising model with three-site interaction on
the Husimi cactus at $i = 2$ (see Fig. 1). As is seen in the
inset in Fig. 1a, the temperature dependencies of the
mapping $f_1(x)$ and, hence, magnetization $M$ in the
region $H \geq 0.1$ (curves specified by Eqs. (\ref{50}) and (\ref{51}) intersect at $H=0.1$), have interesting properties.
When an $H = const$ line intersects only the upper curve (corresponding to Eq. (10) with $i = 1$), the boundaries of
the cyclic window are strictly distinguished (tangent
bifurcation occurs at both edges). This window is represented only by the 3M0 phase (a stable period-3
cycle). As the fixed field decreases, when an $H = const$ line intersects both the upper curve (corresponding to
Eq. (\ref{50}) with $i = 1$) and the lower curve (corresponding to Eq. (\ref{51}) with $i = 1$), a cycle with a period of $3 \times
2^1 = 6$ appears, which corresponds to the modulated
phase 3M1 ${\bf 3}\times2^{\bf 1}=6$) with a period of 6 (see Fig.
2a). The phase transition between the 3M0 and 3M1
phases, accompanied by a change in symmetry, occurs
at the bifurcation points. With a further decrease in the
field, new "bubbles" corresponding to modulated
phases with larger periods will appear on bifurcation
diagrams. Finally, the chaotic regime, which is localized inside the cyclic period-3 window, will be
reached. At the same time, when the magnetization is
considered as a function of the temperature $T$ at a
fixed value $H < 0.1$ or as a function of the magnetic
field $H$ at any fixed temperature, tangent bifurcation occurs only at one edge of the cyclic window \cite{fractals}. A
crisis \cite{crisis, rossler1}, i.e., the collision of the chaotic attractor
with the independent unstable stationary point with a
period of 3, occurs at the other edge (see Fig. 2b). In
this case, modulated phases with a period of $3 \times 2^n$ are
not localized inside the window (similarity with logistic mapping).

\begin{figure*}
\begin{center}
\includegraphics[width=17cm]{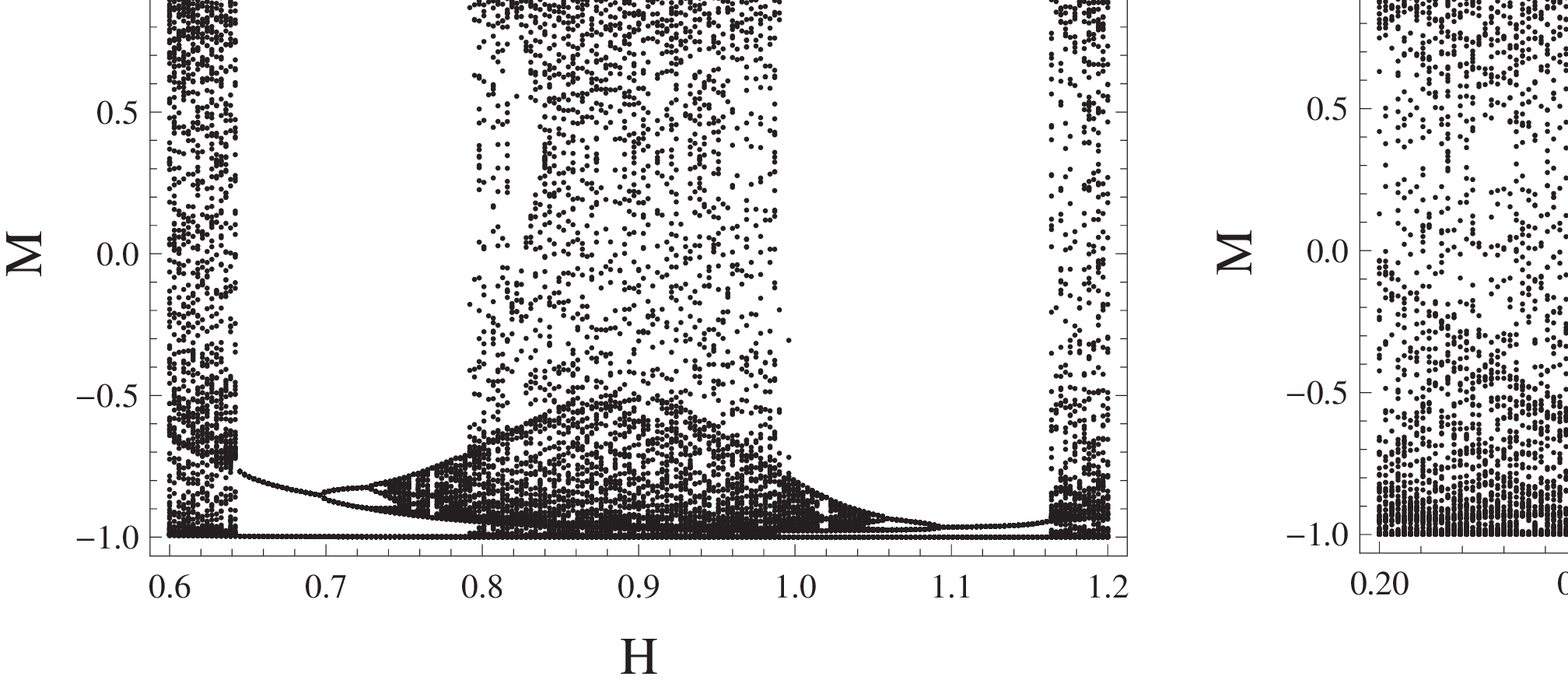}
\caption{\small{Fig. 3. Magnetization in the cyclic period-3 window for
the antiferromagnetic Ising model with three-site interaction at $J_3 = –1$ and $\gamma = 3$ (a) versus the magnetic field for
$T = 0.3$ and (b) versus the temperature for $H = 0.5$.
\label{chain}}}
\end{center}
\end{figure*}

The comparison of the picture described above
with the behavior of the mapping $f_2(x)$ (therefore, the
magnetization in the antiferromagnetic Ising model
with three-site interaction) indicates that tangent
bifurcation occurs at both edges of the window, when
the temperature is fixed ($T = const$) and the external
field is varied (see Fig. 1b). If a $T = const$ line intersects only the curve corresponding to Eq. (\ref{50}) with $i =
2$, the window is represented only by the 3M0 phase.
When a $T = const$ line intersects the curve corresponding to Eq. (\ref{51}) with $i = 2$, there is also the 3M1 phase
(in the form of a new bubble), transition to which
occurs through doubling bifurcation (a phase transition with a change in symmetry). With a further
decrease in the temperature, chaos is reached (as in
the Potts model at $H = const$) inside the window (see Fig. 3a). This picture (localization of $3 \times 2^n$ phases
inside the cyclic period-3 window) for the rational
mappings given by Eqs. (\ref{7}) and (\ref{8.2}), which describe
statistical spin systems, was also observed in the three-dimensional (polynomial) Rossler system \cite{rossler1, rossler}.

At the same time, as is shown in Fig. 3b, the transition between chaos and the 3MO phase through tangent bifurcation in the antiferromagnetic Ising model
with three-site interaction at a fixed field $H$ occurs
only at one edge of the window. At the other edge of
the window, an abrupt change in the chaotic attractor
occurs due to crisis (as in the Potts model at a fixed
temperature $T$). Thus, bifurcation properties in the
cyclic period-3 window in the antiferromagnetic Potts
model on the Bethe lattice in the dependence of the
temperature $T$ at magnetic fields $H$ higher than those
at the intersection point of the curves specified by
Eqs. (\ref{50}) and (\ref{51}) ($H = 0.1$ at $Q = 1.1$) are similar to
the respective properties of the Ising antiferromagnetic model with three-particle interaction on the
Husimi cactus in the dependence of the magnetic field
$H$ (see Fig. 1b and inset in Fig. 1a). However, there is
a certain interval $H\in[1.276; 1.375]$, where the ground state of the Ising antiferromagnetic model
with three-particle interaction is the period-3 phase
3M0 (see Fig. 1b).

Concluding the discussion of the phase structure of
the period-3 window, we consider the Lyapunov exponent $\lambda(x)$ as an order parameter. As is known, $\lambda(x)$
characterizes the degree of the exponential divergence
of two neighboring points induced by the mapping
$x_{n + 1} = f(x_n)$. The exact formula for $\lambda(x)$ has the form

\begin{eqnarray}
\lambda{(x)}=\lim_{n\rightarrow\infty}{\frac{1}{n}}\ln\left|\frac{df^{(n)}(x)}{dx}\right|.\label{21}
\end{eqnarray}

The Lyapunov exponent is negative in the periodic
regime, positive in the chaotic regime, and zero at the
bifurcation points. Figures 4a and 4b show the magnetic field and temperature dependencies of the
Lyapunov exponents for $f_1(x)$ and $f_2(x)$, respectively,
for the same remaining parameters as in Figs. 2a and
3b, respectively. According to Fig. 4, $\lambda(x)=-\infty$ at certain parameters $T$ and $H$. These points correspond to
the superstable cycles \cite{super},which are located in the
region of the particular modulated phase. Therefore,
the construction (both analytical and numerical) of
the superstable cycle of the order $n$ will make it possible to determine the regions of $T$ and $H$ where the
modulated phase of a period of $n$ exists. This problem
will be considered in future works.

\begin{figure*}
\begin{center}
\includegraphics[width=17cm]{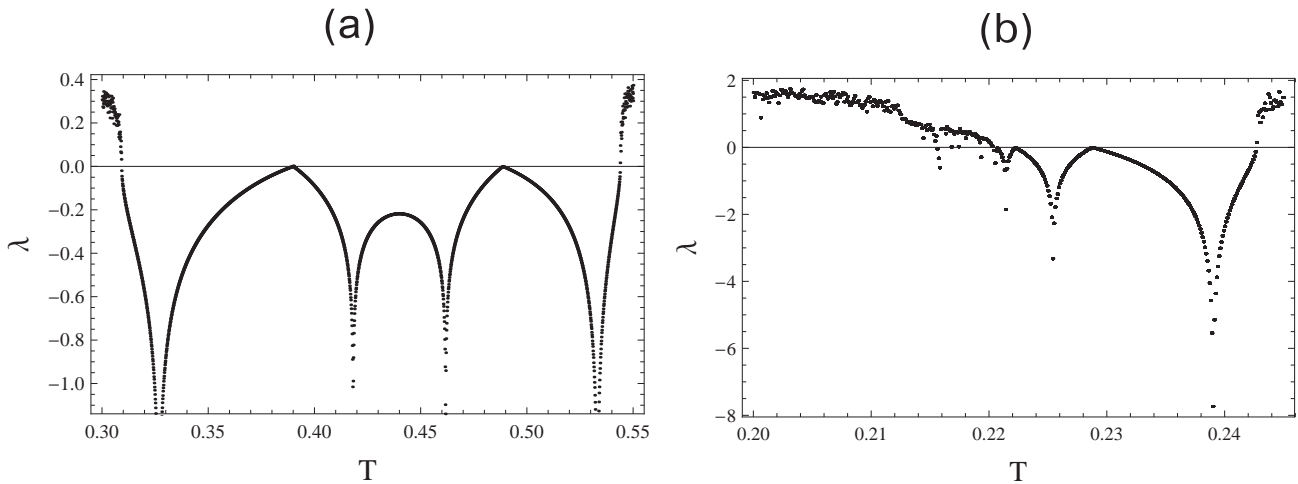}
\caption{\small{Fig. 4. Temperature dependence of the Lyapunov exponent in the cyclic period-3 window for (a) the Potts model
at $Q = 1.1$, $J = –1$, $\gamma = 3$, and $H = 1.24$ and (b) the antiferromagnetic Ising model with three-site interaction at $J_3 =
– 1$, $\gamma = 3$, and $H = 0.5$.
\label{chain}}}
\end{center}
\end{figure*}

Our calculations also show that the Feigenbaum
constants $\alpha$ and $\delta$ for the doubling of the period \cite{Feigenbaum} of
the mappings $f_1(x)$ and $f_2(x)$ converge to the known
universal values $\delta=4,6692...$ and $\alpha=2,5029...$, respectively. Convergence in the period-3 window
(doubling of the period in the form $3 \times 2^n$) is slower
than that in the case of the doubling of the period in
the form $1 \times 2^n$. It is interesting that the universality of
the Feigenbaum constants can also be used for the
approximate construction of the curves of phase transitions between different modulated phases \cite{fractals}.

In summary, a cyclic period-3 window has been
studied in the antiferromagnetic Potts model on the
Bethe lattice and in the antiferromagnetic Ising model
with three-site interaction on the Husimi cactus. The
Bethe lattice and its generalizations are approximations for the standard lattices (Bethe-Peierls approximation), which is much more accurate than the mean field approximation. For rational mappings, which
describe real statistical models (the antiferromagnetic
$Q$-state Potts model on the Bethe lattice and the antiferromagnetic Ising model with three-site interaction
on the Husimi cactus), we have analyzed the mechanism of the transition from the chaotic regime to the
cyclic period-3 window through tangent bifurcation
followed by the doubling cascade $3 \times 2^n$ ($n=2, 3, ...$). The period-3 modulated phase of both models has
been presented on the phase diagram. The Lyapunov
exponents in the period-3 window have been calculated.

This work was supported by the Armenian National
Foundation of Science and Advanced Technologies,
project no. ECSP-09-08 SASP, and the Armenian
National Science and Education Fund, project
no. PS-2497.

\end{document}